\documentclass[preprint,showpacs,preprintnumbers,amsmath,amssymb]{revtex4}

\usepackage{amssymb}

\begin{document}
\def\eqn#1{Eq.$\,$#1}
\def\mb#1{\setbox0=\hbox{$#1$}\kern-.025em\copy0\kern-\wd0
\kern-0.05em\copy0\kern-\wd0\kern-.025em\raise.0233em\box0}
\preprint{}

\title{Statistical mechanics of geophysical turbulence \\
Application to jovian flows and Jupiter's great red spot}
\author{Pierre-Henri Chavanis} 
\affiliation{ Laboratoire de Physique Th\'eorique,  Universit\'e
Paul Sabatier, 118 route de Narbonne, 
31062 Toulouse Cedex 4, France}

\begin{abstract}

We propose a parametrization of two-dimensional geophysical turbulence
in the form of a relaxation equation similar to a generalized
Fokker-Planck equation (Chavanis 2003). This equation conserves
circulation and energy and increases a generalized entropy functional
determined by a prior vorticity distribution fixed by small-scale
forcing (Ellis et al. 2002). We discuss applications of this formalism
to jovian atmosphere and Jupiter's great red spot. We show that, in
the limit of small Rossby radius where the interaction becomes
short-range, our relaxation equation becomes similar to the
Cahn-Hilliard equation describing phase ordering kinetics. This
strengthens the analogy between the jet structure of the great red
spot and a ``domain wall''. Our relaxation equation can also serve as
a numerical algorithm to construct arbitrary nonlinearly dynamically
stable stationary solutions of the 2D Euler equation. These solutions
can represent jets and vortices that emerge in 2D turbulent flows as a
result of violent relaxation. Due to incomplete relaxation, the
statistical prediction may fail and the system can settle on a
stationary solution of the 2D Euler equation which is not the most
mixed state. In that case, it can be useful to construct more general
nonlinearly dynamically stable stationary solutions of the 2D Euler
equation in an attempt to reproduce observed phenomena.

\vskip1cm
\noindent Corresponding author: P.H. Chavanis; e-mail: chavanis@irsamc.ups-tlse.fr; Tel: +33-5-61558231; Fax: +33-5-61556065

\end{abstract}

\maketitle

\newpage

\section{Introduction}
\label{sec_introduction}

An important problem in geophysical fluid turbulence is to obtain a
small-scale parametrization of the 2D Euler equation in order to
compute only large-scale motion. Indeed, in the limit of vanishing
molecular viscosity (which is a limit relevant for geophysical flows),
the 2D Euler equation develops a complicated filamentation at smaller
and smaller scales. This is a redoutable problem for numerical
simulations since those filaments rapidly reach the mesh size and
generate numerical instabilities. Therefore, one usually adds an
``artificial'' or ``numerical'' viscosity $\nu_{T}>0$ in the Euler
equation in order to smooth-out the small scales and regularize the
problem. This viscosity is a simple parametrization of the turbulent
Reynolds tensor which accounts for the correlations of the velocity
fluctuations. However, this artificial viscosity breaks the
conservation laws of the inviscid dynamics, in particular the
conservation of energy, and leads ultimately to a state of rest while
it is known that 2D flows generate long-lived coherent structures.
One can in principle limit these problems by using a small viscosity but 
this is computationaly costly because we have to solve the small-scales in order to solve the large scales which are of physical interest.

How can we improve the Reynolds parametrization and what should we add
to diffusion in order to restore the conservation of energy and obtain
stable stationary states on the coarse-grained scale? It is
illuminating to realize that this problematic is related to the one
encountered by Einstein in his investigation of Brownian motion and by
Chandrasekhar in his stochastic approach of stellar dynamics
\cite{chandra}. The solution is now well-known: they solved the
problem by introducing a dynamical friction in addition to diffusion
so as to recover the correct equilibrium state.  In their cases, the
equilibrium state is the Maxwell-Boltzmann distribution predicted by
statistical mechanics. The condition that the Maxwell-Boltzmann
distribution is the stationary solution of the Fokker-Planck equation
leads to the celebrated Einstein relation $\xi=D\beta$ between the
coefficient of diffusion and the friction. Can we follow a similar
approach in 2D turbulence and which adaptations should we make?

Two theories have been developed to predict the (meta)-equilibrium state
resulting from violent relaxation in terms of statistical
mechanics. The approach of Miller \cite{miller} and Robert \& Sommeria
\cite{rs} applies to unforced systems which are strictly described by
the 2D Euler equation.  It predicts the most probable distribution of
the flow depending on the energy, the circulation and on the infinite
set of Casimirs (or vorticity moments) that are conserved by the Euler
equation. This approach is quite powerful in the case of simple
initial conditions made of a small number of patches for which the
number of control parameters is reduced. It may also describe more
general situations in the limit of strong mixing or low energy studied
by Chavanis \& Sommeria
\cite{jfm1} which is characterized by a finite number of relevant
contraints (the first moments of vorticity). However, the approach of
\cite{miller,rs,jfm1} also presents limitations that have been
discussed by several authors. In particular, the statistical
prediction presupposes that the (fine-grained) initial conditions are
known. Indeed, the equilibrium state depends not only on circulation
(the equivalent of mass) and energy as in ordinary statistical
mechanics but also on the values of all the vorticity moments
$\Gamma_{n}=\int
\overline{\omega^{n}}d^{2}{\bf r}$ which cannot be determined 
from the coarse-grained field when the vorticity has mixed (for $t>0$)
since $\overline{\omega^{n}}\neq
\overline{\omega}^{n}$. Since the initial conditions are not known in
general (and we are never sure whether a flow is already mixed or
not), we cannot really make statistical {predictions}. This approach
also assumes that the evolution is inviscid and freely
evolving. However, in real situations this is not the case and the
conservation of all the Casimirs is abusive.

An alternative approach has been developped by Turkington and
collaborators \cite{turk,ellis} in order to remedy these
difficulties. This approach is particularly relevant for geophysical
and astrophysical flows that are forced at small-scales. For example,
convective plumes are known to form regularly in the atmosphere of
Jupiter. Their significant horizontal divergence coupled with rotation
leads to small-scale anticyclonic vorticity production. Therefore,
there is a permanent forcing and dissipation at small scales. These
effects destroy the conservation of the Casimirs while establishing
instead a global distribution of vorticity. In that case, it may be
more relevant to fix a {\it prior} distribution of vorticity
$\chi(\sigma)$ instead of the Casimirs. This amounts to treating the
constraints associated with the fragile integrals $\Gamma_{n\ge 2}$
 canonically while the robust integrals $\Gamma$ and $E$  are treated
microcanonically. In this point of view, we abandon the strict
conservation of all the Casimirs and keep their influence indirectly
in the function $\chi(\sigma)$. This prior distribution has to be
adapted to the geophysical situation as discussed in \cite{ellis}.
This is similar to our idea \cite{gfp} of fixing a generalized entropy
$S=-\int C(q)d^{2}{\bf r}$ a priori, depending on the situation
contemplated. In fact, we will show that the prior vorticity
distribution $\chi(\sigma)$ directly determines a form of generalized
entropy $C(q)$.

Once a prior vorticity distribution is given, the generalized entropy
is known and the equilibrium state depends only on the value of
circulation and energy (robust constraints). We are thus in a
situation of usual thermodynamics but with a generalized form of
entropy. It is therefore possible to apply phenomenological arguments
similar to those given by Einstein and Chandrasekhar to derive a
relaxation equation towards the equilibrium state which conserves
energy and circulation and increases the generalized entropy. This
equation has the form of a generalized Fokker-Planck equation
\cite{gfp}  involving a turbulent viscosity (diffusion) and a drift. As in Brownian theory, the diffusion and the drift
coefficients are connected by a sort of Einstein relation involving
here a negative temperature.

In this paper, after a brief summary of the two statistical theories
of 2D turbulence (Sec. \ref{sec_sum}), we give in Sec. \ref{sec_para}
a more precise construction and physical justification of the
relaxation equation introduced in \cite{gfp}, emphasizing the
connexion with the work of Ellis {\it et al.} \cite{ellis}. We present
a new derivation of this generalized Fokker-Planck equation by using
an analogy with Onsager's linear thermodynamics and with the theory of
Brownian motion. The domain of application of our parametrization is
therefore better established and the meaning of the generalized
entropy, determined by the prior vorticity distribution, clearer. For
given prior vorticity distribution $\chi(\sigma)$, there is no free
parameter in our model. In Sec. \ref{sec_small}, we apply this
formalism to the atmosphere of Jupiter. We consider the limit of small
Rossby deformation radius, corresponding to short-range interactions,
and show that our relaxation equation reduces to a form of
Cahn-Hilliard equation that is used in the theory of phase ordering
kinetics
\cite{bray}. Its stationary solutions describe ``domain walls''
accounting for the annular jet structure of Jupiter's great red
spot. On the other hand, according to the refined stability theorem of
Ellis {\it et al.} \cite{ellis}, our relaxation equation can also be
used as a numerical algorithm to construct arbitrary nonlinearly
dynamically stable stationary solutions of the 2D Euler equation. This
can be used to construct confined structures that are not described by
the statistical theory because of incomplete relaxation (lack of
mixing/ergodicity) \cite{brands}. This paper discusses the main ideas
and presents the formalism. Numerical simulations illustrating these
results will be given in a future work.

\section{Summary of the statistical theories}
\label{sec_sum}

\subsection{The quasi-geostrophic equations}
\label{sec_qg}

Large-scale oceanic and atmospheric flows are quasi two-dimensional
due to planetary rotation and stratification \cite{pedlosky}. This
property prevents vortex stretching and can lead to the formation of
large-scale coherent structures (e.g., Jupiter's great red spot,
cyclonic and anticyclonic atmospheric eddies,...) with robust
stability (Jupiter's great red spot was observed more than three
centuries ago). In addition, in geophysical situations, the Reynolds
numbers are so high that viscosity is not expected to play a crucial
role for the timescales and length scales of interest. A relevant
model is provided by the quasi-geostrophic (Q.G.) equations
\begin{equation}
{\partial q\over\partial t}+{\bf u}\cdot \nabla q=0,
\label{e1}
\end{equation}
\begin{equation}
q=-\Delta\psi+{1\over R^{2}}\psi-h(y),
 \label{e2}
\end{equation}
where $q$ is the potential vorticity (PV), $\psi$ the stream-function defined by ${\bf u}=-{\bf z}\times\nabla\psi$,  $R$ is the Rossby radius
of deformation and $h(y)$ is an effective topography including the
$\beta$-effect. The second term in Eq. (\ref{e2}) is similar to the
Debye shielding in plasma physics. More generally, we can consider a
relation of the form $\psi({\bf r},t)=\int u(|{\bf r}-{\bf r}'|)
\lbrack q({\bf r}',t)+h(y')\rbrack d^{2}{\bf r}'$ where $u(|{\bf r}-{\bf r}'|)$ is a binary potential of interaction. In the Q.G. model, one has
$u(\xi)=(1/2\pi)K_0(\xi/R)$, where $K_0(z)$ is a modified Bessel
function. It has the asymptotic behaviours: $K_{0}(z)\sim -\ln z$ for
$z\rightarrow 0$ and $K_{0}(z)\sim \sqrt{\pi\over 2z}e^{-z}$ for
$z\rightarrow +\infty$. Stationary solutions of the Q.G. equations are
specified by a relation of the type $q=f(\psi)$ where $f$ is an
arbitrary function.

The Q.G. equation (\ref{e1}) simply states that, in the
absence of viscosity, the potential vorticity $q$ is conserved by the
flow. The Q.G. equations conserve the energy
\begin{equation}
E={1\over 2}\int (q+h)\psi \ d^{2}{\bf r},
\label{e3}
\end{equation}
and the Casimirs $C_{h}=\int h(q)d^{2}{\bf r}$, where $h$ is
any continuous function of $q$. This infinite set of
constraints is due to the inviscid dynamics and results
from the conservation of $q$ and the incompressibility of the
flow. The conservation of the Casimirs is equivalent to the
conservation of the PV moments
\begin{equation}
\Gamma_{n}=\int q^{n}\ d^{2}{\bf r}, \label{e4}
\end{equation}
which include the circulation $\Gamma$ and the enstrophy
$\Gamma_{2}$. The moments $\Gamma_{n}$ with $n\ge 2$ will be called
fragile constraints because they are affected by viscosity or
coarse-graining (since $\overline{q^{n}}\neq
\overline{q}^{n}$) while the circulation $\Gamma$ and energy $E$ will
be called robust constraints because they are approximately conserved
on the coarse-grained scale or in the presence of a small viscosity.

\subsection{The statistical equilibrium state}
\label{sec_vr}

The Q.G. equations generate a complicated {mixing process} and
develop filaments at smaller and smaller scales. If
we work with a finite resolution $\epsilon$, the coarse-grained
PV $\overline{q}({\bf r},t)$ (the local average of
$q({\bf r},t)$ over a cell of size $\epsilon^{2}$) will reach a
{\it metaequilibrium} state on a very short timescale. This is an
inviscid violent relaxation driven by purely inertial effects
(mixing). In practice, a natural coarse-graining is played by inherent
viscosity which smoothes out the small scales. The resulting structure
has the form of a large-scale vortex or a jet.

There has been some attempts to describe this metaequilibrium state in
terms of statistical mechanics \cite{miller,rs,jfm1}.  The same ideas
had been developed previously by Lynden-Bell \cite{lb} in stellar
dynamics for the Vlasov-Poisson system (see \cite{csr,houches} for a
description of this analogy). Assuming that mixing is efficient, which
is linked to a condition of ergodicity, the most probable local
distribution of vorticity $\rho({\bf r},\sigma)$ is obtained by
maximizing the mixing entropy
\begin{equation}
\label{e9}
S\lbrack \rho\rbrack=-\int \rho\ln\rho \ d^{2}{\bf r}d\sigma,
\end{equation}
taking into account all the constraints imposed by the dynamics.
This leads to the Gibbs state
\begin{equation}
\label{e10}
\rho({\bf r},\sigma)={1\over Z({\bf r})}\chi(\sigma)  e^{-(\beta\psi+\alpha)\sigma},
\end{equation}
where $\chi(\sigma)\equiv {\rm
exp}(-\sum_{n>1}\alpha_{n}\sigma^{n})$ accounts for the
conservation of the fragile moments $\Gamma_{n>1}=\int
\rho\sigma^{n}d^{2}{\bf r}d\sigma$ and the ``partition function''
$Z=\int\chi(\sigma)e^{-(\beta\psi+\alpha)\sigma}d\sigma$ is determined
by the local normalization condition $\int\rho d\sigma=1$ ($\alpha$
and $\beta$ are the usual Lagrange multipliers associated with the
robust constraints $\Gamma$ and $E$). 
Using the Gibbs state (\ref{e10}),  the coarse-grained PV
$\overline{q}=\int
\rho\sigma d\sigma$ is related to the streamfunction according to
\begin{equation}
\label{e11} \overline{q}=-{1\over\beta}{\partial\ln
Z\over\partial\psi}=F(\beta\psi+\alpha)=f(\psi).
\end{equation}
Thus, for given initial conditions, the statistical theory selects a
particular stationary solution of the Q.G. equations (on the
coarse-grained scale) among all possible ones.  Specifically, the
equilibrium state is obtained by solving the differential equation
\begin{equation}
\label{e11posson} -\Delta\psi+{1\over R^{2}}\psi-h(y)=f_{\alpha_{n},\beta}(\psi),
\end{equation}
and relating the Lagrange multipliers $\alpha_{n}$, $\beta$ to the
constraints $\Gamma_{n}$, $E$. In this approach, the function
$\chi(\sigma)$ is determined from the constraints {\it a
posteriori}. Indeed, we need to solve the full problem in order to get
the expression of $\chi(\sigma)$. In this sense, the vorticity moments
are treated microcanonically.  Taking the derivative of Eq.
(\ref{e11}), it is easy to show that
\begin{equation}
\label{e12}
\overline{q}'(\psi)=-\beta q_{2}, \qquad q_{2}\equiv \int \rho (\sigma-\overline{q})^{2}
d\sigma\ge 0,
\end{equation}
where $q_{2}$ is the local centered enstrophy. Therefore,
$f(\psi)$ is a monotonic function increasing at negative temperatures
and decreasing at positive temperatures.

\subsection{Prior vorticity distribution and generalized entropy}
\label{sec_ge}

For freely evolving 2D turbulence, the function $\chi(\sigma)$ is
determined by the initial conditions through the values of the Casimir
integrals. However, in geophysics, there exists situations in which
the flow is continuously forced at small-scales so that the
conservation of the Casimirs is destroyed.  Ellis {\it et al.}
\cite{ellis} have proposed to take into account these situations by
treating the function $\chi(\sigma)$ canonically (see Appendix
\ref{sec_relat}).  Therefore, the function $\chi(\sigma)$ must be
viewed as a prior vorticity distribution fixed by the small-scale
forcing. Its specific form has to be adapted to the situation.  In
that context, the local vorticity distribution can be written
$\rho({\bf r},\sigma)=\chi(\sigma)\mu({\bf r},\sigma)$ where
$\chi(\sigma)$ is {\it given}. The optimal distribution is then
obtained by maximizing the relative entropy \cite{ellis}:
\begin{equation}
\label{ngr}
S[\rho]=-\int \mu({\bf r},\sigma)\ln \mu({\bf
r},\sigma) \chi(\sigma) d^{2}{\bf r}d\sigma,
\end{equation}
at fixed mass, energy and normalization (no other constraint). This leads to
Eq. (\ref{e10}) but with a different interpretation. The local
distribution of vorticity $\rho({\bf r},\sigma)$ is equal to the
global vorticity distribution $\chi(\sigma)$ locally modified by the
large-scale flow through the Boltzmann factor ${1\over
Z({\bf r})}e^{-(\beta\psi({\bf r})+\alpha)\sigma}$.  The coarse-grained
vorticity is still given by Eq. (\ref{e11}). However, the crucial
point to realize is that the function $F$ is now entirely determined
by the prior vorticity distribution $\chi(\sigma)$ while, in the
former case, it had to be related to the initial conditions in a very
complicated way. This makes this new approach very attractive. The resulting 
macro-distribution $\overline{q}(\epsilon)={1\over Z(\epsilon)}\int \chi(\sigma)e^{-\sigma\epsilon}d\sigma$ with $\epsilon=\beta\psi+\alpha$ is formally similar to the notion of 
superstatistics developed by Beck \& Cohen \cite{bc}, see
\cite{super} for the description of this analogy.

We now show that the prior distribution function determines a form of
generalized entropy. Since the relation
$\overline{q}=F(\beta\psi+\alpha)$ is monotonic, it can be obtained by
maximizing a functional of the form
\begin{equation}
S=-\int C(q)d^{2}{\bf r},
\label{se6}
\end{equation} 
where $C$ is a convex function (i.e. $C''>0$), at fixed circulation
and energy. Indeed, introducing appropriate Lagrange multipliers and
writing the variational principle in the form
\begin{equation}
\label{e6} \delta S-\beta\delta E-\alpha\delta \Gamma=0,
\end{equation}
we find that the critical points of $S$ at
fixed circulation and energy are given by
\begin{equation}
\label{e7} C'(q)=-\beta\psi-\alpha.
\end{equation}
Since $C'$ is a monotonically increasing function of $q$, we can inverse 
this relation to obtain
\begin{equation}
\label{e7b} q=F(\beta \psi+\alpha)=f(\psi),
\end{equation}
where 
\begin{equation}
\label{e7c} F(x)=(C')^{-1}(-x).
\end{equation}
From the identity
\begin{equation}
\label{e8} q'(\psi)=-\beta/C''(q),
\end{equation}
resulting from Eq. (\ref{e7}), $q(\psi)$ is monotonically decreasing
if $\beta>0$ and monotonically increasing if $\beta<0$. Thus
Eq. (\ref{e7b}) is equivalent to Eq. (\ref{e11}) provided that we make
the identification (\ref{e7c}). Therefore, for any function $F(x)$ determined
by the prior vorticity distribution $\chi(\sigma)$, one can associate
a generalized entropy (\ref{se6}) where $C(q)$ is given by Eq. (\ref{e7c}) or
equivalently
\begin{equation}
\label{e8b} C(q)=-\int^{q}F^{-1}(x)dx.
\end{equation}
Furthermore, $S[\overline{q}]$ really has the status of an
entropy in the sense of large deviations. Indeed, Ellis {\it et al.} 
\cite{ellis} show that the distribution probability of the
coarse-grained vorticity field can be written in the form of Cramer's
formula $P[\overline{q}]\sim e^{nS[\overline{q}]}$ where $n$
is the number of sites of the underlying lattice introduced in their
mathematical analysis.  Therefore, the optimal PV field
$\overline{q}$ maximizes $S[\overline{q}]$ at fixed
circulation and energy. They also relate the entropy
$S[\overline{q}]$ to the prior distribution $\chi(\sigma)$ by
using the formalism of large deviations. The steps
(\ref{se6})-(\ref{e8b}) presented above are a simple alternative to
construct the generalized entropy associated with $\chi(\sigma)$. It
of course leads to the same result as \cite{ellis}.

\subsection{Refined condition of nonlinear dynamical stability}
\label{sec_refined}

Ellis {\it et al.} \cite{ellis} also show that the microcanonical
variational principle
\begin{equation}
\label{r1} {\rm Max}\quad S[q]\quad |\quad E[q]=E,\quad \Gamma[q]=\Gamma,
\end{equation}
provides a refined condition of
nonlinear dynamical stability for the Q.G. equations.  
It is stronger than the canonical variational principle
\begin{equation}
\label{r2} {\rm Max}\quad J[q]=S[q]-\beta E[q]\quad |\quad \Gamma[q]=\Gamma,
\end{equation}
which just provides a sufficient condition of nonlinear dynamical
stability ($J$ is equal to the free energy $F=E-TS$ multiplied by
$-\beta$ where $\beta=1/T$ is usually negative). The two variational
principles differ if the ensembles are not equivalent which is the
case for systems with long-range interactions. This can have important
implications for geophysical flows since vortex solutions can be
stable in the microcanonical ensemble while they are unstable in the
canonical ensemble \cite{ellis,turk}. This is similar to 
``core-halo'' structures with negative specific heats in astrophysics
\cite{fermi}.  Note that, in the view point of Ellis {\it et al.}, the
statistical equilibrium macrostates are stable with respect to the
perturbations on the microscopic scale (thermodynamical stability) and
the steady mean state $\overline{q}$ is stable with respect to
macroscopic perturbations (nonlinear dynamical stability).

It is worth mentioning, however, that the results of Ellis {\it et al.} 
\cite{ellis} on nonlinear dynamical stability are valid 
independently of the statistical theory. They
apply to any stationary solution of the Q.G. equation, whether or not
it corresponds to the statistical equilibrium state (\ref{e11}). This
is important in the case of incomplete relaxation where the
statistical prediction fails \cite{brands}. The stability results
still apply but in that case $S$ is a H-function \cite{tremaine,gfp},
not an entropy and $J$ is an Energy-Casimir functional \cite{holm},
not a free energy. Furthermore, the variational principles (\ref{r1})
and (\ref{r2}) have not a direct thermodynamical interpretation in
that case. However, because of the formal ressemblance of these
variational principles with those of thermodynamics, it is possible to
develop a {\it thermodynamical analogy}
\cite{gfp} to settle the nonlinear dynamical stability of a stationary
solution of the Q.G. equations. In this context, the ``canonical''
stability criterion (\ref{r2}) is equivalent to the criterion of
formal nonlinear stability given by Holm {\it et al.} \cite{holm}. The
Arnold's theorems provide particular conditions of formal nonlinear
stability in the sense of (\ref{r2}). The ``microcanonical'' stability
criterion (\ref{r1}) proved by Ellis {\it et al.}
\cite{ellis} refines these theorems. In particular, a flow can be
stable by the criterion (\ref{r1}) although it violates the Arnold's theorems.

We note finally that the stability criterion (\ref{r1}) is consistent
with the phenomenology of 2D turbulence. Indeed, the H-functions
increase on the coarse-grained scale (in the sense that
$S[\overline{q}(t)]\ge S[\overline{q}(0)]$ where
$\overline{q}(0)=q(0)$ is un-mixed \cite{tremaine}) or in the presence
of a small viscosity (since $\dot S=\nu\int C''(q)(\nabla
q)^{2}d^{2}{\bf r}\ge 0$), while the energy and the circulation are
approximately conserved. Therefore, the metaequilibrium state is
expected to maximize a certain $H$-function at fixed circulation and
energy.  This generalizes the concept of ``selective decay'' which
considers the decrease of enstrophy at fixed energy and
circulation. However, the 2D Euler equation does not select a
universal $H$-function. Minus the enstrophy $-\Gamma_{2}=-\int
q^{2}d^{2}{\bf r}$ and Tsallis $Q$-functionals $S_{Q}=-{1\over
Q-1}\int (q^{Q}-q)d^{2}{\bf r}$ are {\it particular} $H$-functions
(not entropies) that {sometimes} occur in 2D turbulence in case of
incomplete relaxation as in the plasma experiment of Huang
\& Driscoll \cite{hd}. They are associated with confined stationary 
solutions of the 2D Euler equation that have a compact support
(``polytropic vortices'') \cite{gfp}. Note that many other confined
solutions can emerge in 2D turbulence, e.g. \cite{staquet}, which are
not described by Tsallis of by the enstrophy functionals.

\section{A parametrization of geophysical flows}
\label{sec_para}

In the context of freely evolving 2D turbulence, a thermodynamical
parametrization of the 2D Euler equation has been proposed by Robert
\& Sommeria \cite{rsmepp} in terms of relaxation equations based on a Maximum Entropy Production Principle.  These equations conserve all the 
Casimirs, increase the mixing entropy (\ref{e9}) and relax towards the
Gibbs state (\ref{e10}). In the situation considered by Ellis {\it et
al.} \cite{ellis} where the system is forced at small-scales, we can
propose an alternative parametrization of the 2D Euler equation. In
that case, we have seen that only the energy and the circulations are
conserved. The conservation of the Casimirs is replaced by a prior
vorticity distribution $\chi(\sigma)$. This fixes the form of
generalized entropy. We can thus propose to describe the large-scale
evolution of the flow in that case by a relaxation equation which
conserves energy and circulation and increases the generalized entropy
(\ref{se6}) associated with $\chi(\sigma)$.

Taking the local average of the Q.G. equation (\ref{e1}), we have
\begin{equation}
\label{ax0}
{\partial \overline{q}\over\partial t}+{\bf u}\cdot \nabla \overline{q}=
-\nabla\cdot {\bf J},
\end{equation}
where ${\bf J}=\overline{\tilde q\tilde {\bf u}}$ is the Reynold's diffusion
current.  We shall determine its expression by using a procedure
similar to Onsager's linear thermodynamics.  Noting that the
``chemical potential'' $\alpha({\bf r},t)=-\beta\psi-C'(q)$ is uniform
at equilibrium according to Eq. (\ref{e7}), we take ${\bf
J}=D'\nabla\alpha$. This choice yields a parametrization of the form
\begin{equation}
\label{ax1}
{\partial {q}\over\partial t}+{\bf u}\cdot \nabla {q}=\nabla\cdot \biggl\lbrace D\biggl\lbrack \nabla{q}+{\beta\over C''({q})}\nabla\psi\biggr\rbrack\biggr\rbrace,
\end{equation}
where, for convenience, we have droped the bars and set
$D=D'C''(q)$. The convex function $C(q)$ is a {\it non-universal}
function which encapsulates the complexity of the fine-grained
dynamics and depends on the situation contemplated. It is determined
by the prior distribution $\chi(\sigma)$ which encodes the statistics
of vorticity fluctuations induced by small-scale forcing
\cite{ellis}. Equation (\ref{ax1}) can be interpreted as a 
generalized Fokker-Planck equation \cite{gfp}. We note that the term
arising in the eddy flux in addition to the usual diffusion is a {\it
systematic drift} $-\xi\nabla\psi$. This drift {\it must} exist in
order to yield the correct distribution at equilibrium. The necessity
of this drift is similar to the necessity of the dynamical friction in
stellar dynamics (see the first, phenomenological, approach of
Chandrasekhar \cite{chandra} on dynamical friction).  The drift
coefficient (mobility) is given by $\xi=D\beta$ which we link to the
celebrated Einstein relation. Thus, the systematic drift is the
counterpart of the ordinary {\it dynamical friction} and the preceding
relation is the counterpart of the ``fluctuation-dissipation'' theorem
in Brownian theory. This drift is usually absent from usual modelling
of 2D turbulence where only a viscosity (or hyperviscosity) is
introduced. This corresponds to the infinite temperature limit
($\beta=0)$ of our parametrization. The physical origin of the drift
should be understood from kinetic theories as attempted in \cite{prl}
using a quasilinear theory of the Euler equation or in \cite{kin} for
point vortices. Unfortunately, these formal methods do not describe
the very nonlinear phase of ``violent relaxation'' so that more
phenomenological methods are prefered in that case.

In Onsager's linear thermodynamics, the inverse temperature $\beta$ is
constant. This is appropriate to a {\it canonical} situation. In that
case, the generalized Fokker-Planck equation (\ref{ax1}) increases the
free energy $J=S-\beta E$ at fixed circulation provided that the
diffusion coefficient $D$ is positive. However, statistical ensembles
can be inequivalent in 2D turbulence \cite{ellis} and the correct
description is the {\it microcanonical} ensemble.  The conservation of
energy (\ref{e3}) can be taken into account by determining the
evolution of $\beta(t)$ in Eq. (\ref{ax1}) from the requirement $\dot
E=0$ as in
\cite{rsmepp}. This yields
\begin{equation}
\label{ax2}
\beta(t)=-{\int D\nabla{q}\cdot \nabla\psi d^{2}{\bf r}\over \int D{(\nabla\psi)^{2}\over C''({q})}d^{2}{\bf r}}.
\end{equation}
With this modification, the generalized Fokker-Planck equation
(\ref{ax1}) increases the generalized entropy $S$ at fixed circulation
and energy until the maximum entropy state is reached. It
automatically determines the right values of the Lagrange multipliers
$\alpha$ and $\beta=\lim_{t\rightarrow +\infty}\beta(t)$ at
equilibrium. This relaxation equation selects only maxima of entropy,
not minima or saddle points that are linearly unstable
\cite{gfp}. Note that the relaxation equation (\ref{ax1}) with
(\ref{ax2}) can also be obtained from a Maximum Entropy Production
Principle, see \cite{gfp}, by using a procedure similar to that
employed by Robert \& Sommeria \cite{rsmepp} in their parametrization.

Heuristic arguments \cite{rr,csr} or more formal kinetic theory
\cite{prl} suggest that the turbulent diffusion in Eq. (\ref{ax1}) is
given by $D=K\epsilon^{2}q_{2}^{1/2}$ where $K$ is a constant of order unity,
$\epsilon$ is the scale of unresolved fluctuations and $q_{2}$ the
local centered enstrophy (see Appendix \ref{sec_diffcoeff}). This space dependant diffusion coefficient, related to the fluctuations of vorticity,
can account for {\it incomplete relaxation}
\cite{rr} and was used in oceanic modelling \cite{kazantsev}.  Now,
from Eqs. (\ref{e8}) and (\ref{e12}) we have the important relation
\begin{equation}
\label{org}
q_{2}={1\over C''(q)},
\end{equation}
which can be seen as a sort of fluctuation-dissipation theorem
\cite{shallow}.  This equation is valid at equilibrium but, in the spirit
of linear thermodynamics, we shall still use it out of equilibrium as
a convenient approximation (see also Appendix \ref{sec_noneq}). In
that case, the diffusion coefficient can be expressed in terms of $q$
alone as
\begin{equation}
\label{e18}
D={K\epsilon^{2}\over \sqrt{C''(q)}}.
\end{equation}
Note that we can also obtain the parametrization (\ref{ax1}) by
closing the hierarchy of moment equations of \cite{rsmepp} with
relation (\ref{org}), see \cite{gfp}. In this sense, (\ref{org}) can
be regarded as a closure relationship.

We have thus obtained a parametrization of 2D geophysical flows with
no free parameter once the prior distribution of vorticity is fixed.
These equations are expected to be valid close to the equilibrium
state in the spirit of Onsager's linear thermodynamics. However, they
may offer a useful parametrization even if we are far from
equilibrium. Alternatively, according to the refined stability
criterion of Ellis {\it et al.} \cite{ellis}, the relaxation equation
(\ref{ax1}) can also provide a poweful numerical algorithm to compute
arbitrary nonlinearly dynamically stable stationary solutions of the
Q.G. equations characterized by a function $q=f(\psi)$ where $f$ is
related to $C$ by Eq. (\ref{e7c}). This can be used to construct
confined solutions that are not consistent with statistical theory
because of incomplete relaxation. We need just start from an initial
condition (guess) with given $(E,\Gamma)$ and
Eqs. (\ref{ax1})-(\ref{ax2}) will relax towards a stationary solution
specified by $f$ or $C$ which is, by construction, a nonlinearly
dynamically stable stationary solution of the Q.G. equations with
these prescribed values of energy and circulation. In that context,
since we are only interested by the equilibrium state, we can take
$D={\rm Cst.}$ and even drop the advective term in
Eq. (\ref{ax1}). Another numerical algorithm has been proposed by
Turkington \& Whitaker
\cite{tw} based on iterative schemes techniques and it was used with
success in many situations. The relaxation equations
(\ref{ax1})-(\ref{ax2}) can offer an alternative numerical algorithm.
We again emphasize that constructing exact, nonlinearly dynamically
stable stationary solutions of the 2D Euler equation is an important
problem in itself and is an interest of the relaxation equation
(\ref{ax1}), independently of the statistical theory. In particular,
it is difficult to solve the differential equation (\ref{e11posson})
directly (for flows without simple symmetries) and be sure that the
solution is stable.

Some examples of prior vorticity distributions and corresponding
generalized entropies and parametrizations have been collected in
\cite{gfp,super}. For example, in the two-levels case, the 
generalized entropy is similar
to the Fermi-Dirac entropy
\begin{equation}
\label{e18w}
S[q]=-\int \biggl\lbrack p\ln p+(1-p)\ln (1-p)\biggr\rbrack d^{2}{\bf r},
\end{equation}
where $\overline{q}=p\sigma_{1}+(1-p)\sigma_{0}$.  We note that, in
this particular case, the generalized entropy $S[q]$ coincides with
the mixing entropy $S[\rho]$ defined in (\ref{e9}) since $\rho({\bf
r},\sigma)= p({\bf r})\delta (\sigma-\sigma_{1}) +(1-p({\bf r}))
\delta (\sigma-\sigma_{0})$ where $p$ is expressed in terms of $q$. It
also coincides with the relative entropy $S[\rho]$ defined in (\ref{ngr}) since
$\chi(\sigma)=\delta (\sigma-\sigma_{1}) +\chi \delta
(\sigma-\sigma_{0})$.  This implies that the two-levels case can be
seen either as the statistical equilibrium state resulting from the
free merging of large patches with two types of vorticity (point of
view of Sec.
\ref{sec_vr}) {\it or} as a particular
type of prior vorticity distribution corresponding to a small-scale
forcing with two intense peaks of vorticity (point of view of Sec. 
\ref{sec_ge}). Although the equations
are the same, their interpretation is different. We think that the
second interpretation is the most relevant for geophysical flows due
to the existence of convective plumes.  On the other hand, in the
geophysical situation considered by Ellis et al. \cite{ellis}, the
prior vorticity is taken as a decentered gamma function and the generalized
entropy is
\begin{equation}
\label{e18q}
S[q]=-{1\over\lambda Q_{2}}\int\biggl\lbrack q-{1\over\lambda}\ln (1+\lambda q)\biggr\rbrack d^{2}{\bf r},
\end{equation}
where $Q_{2}$ and $2\lambda Q_{2}^{1/2}$ are the variance and the
kurtosis of $\chi(\sigma)$. For $\lambda\rightarrow 0$, $S[q]$ is
proportional to minus the enstrophy $\Gamma_{2}=\int q^{2}d^{2}{\bf
r}$. Note that minus the enstrophy can represent either a H-function
arising in case of incomplete relaxation as in \cite{hd}, a
generalized entropy associated with a gaussian distribution
$\chi(\sigma)$ as in \cite{miller} or the approximate form of mixing entropy
(\ref{e9}) in the limit of strong mixing as in \cite{jfm1}. It can
thus have several interpretations.

These results can be readily extended to the more realistic
shallow-water (SW) equations whose statistical mechanics has been
developed by Chavanis \& Sommeria \cite{shallow} in the case
of unforced flows. To apply the point of view of Ellis {\it et al.} 
\cite{ellis}, we must write the total vorticity distribution as
$\rho({\bf r},\sigma)=\chi(\sigma)\mu({\bf r},\sigma)$ where
$\chi(\sigma)$ is {\it given}. The optimal distribution $\rho^{*}({\bf
r},\sigma)$ is then obtained by maximizing the entropy $S[\rho]=-\int
\mu({\bf r},\sigma)\ln\mu({\bf r},\sigma)\chi(\sigma)h({\bf r}) d^{2}{\bf
r}d\sigma$ ($h$ is the elevation of the fluid) at fixed mass,
circulation and energy (no other constraints). This leads to the same
results as in \cite{shallow}, equivalent to Eq. (\ref{e10}), except that
now the prior vorticity distribution $\chi(\sigma)$ must be regarded
as given. The coarse-grained PV maximizes a generalized entropy
$S[q]=-\int C(q)h d^{2}{\bf r}$ at fixed energy, mass and circulation,
where $C(q)$ is determined by $\chi(\sigma)$ as in
Sec. \ref{sec_ge}. Relaxation equations equivalent to Eq. (\ref{ax1})
are obtained by maximizing $\dot S[q]$ at fixed $E$. This yields the
same relaxation equations as in
\cite{shallow} except that now $q_{2}=1/C''(q)$.

\section{The limit of small Rossby radius }
\label{sec_small}

We shall now discuss some applications of these ideas to geophysical
and astrophysical flows.  The application of the statistical theory of
Miller-Robert-Sommeria \cite{miller,rs} to jovian flows and Jupiter's great red
spot has been developed in \cite{nore,mich,bs,bd}. In particular,
Bouchet \& Sommeria \cite{bs} consider a limit of small deformation
radius $R\rightarrow 0$ and interprete the annular jet structure of
Jupiter's great red spot (GRS) as the coexistence of two
thermodynamical phases in contact, like in the van der Waals phase
transition (an analogy sketched in
\cite{nore}).  They show in particular that a very good agreement with the
GRS can be obtained in the two-levels approximation of the statistical
theory, and that this approximation can be motivated by the existence
of convecting plumes in the jovian atmosphere. Turkington and
collaborators \cite{turk,ellis} have also developed a model of
Jupiter's great red spot based on their statistical mechanics
approach.  In particular, they consider a realistic topography (Limaye
profile) and show that a vortex structure appears precisely at the
latitude of Jupiter's great red spot. They also emphasize the
inequivalence of statistical ensembles and the relevance of a
microcanonical formulation to settle the stability of jets and
vortices.

We shall here discuss some connexion between the relaxation equation
(\ref{ax1}) and the Cahn-Hilliard equation of phase ordering
kinetics. This connexion will strengthen the resemblance of Jupiter's
great red spot with a ``domain wall''.  The relaxation equation
(\ref{ax1}) can be written
\begin{equation}
\label{ne15} {\partial {q}\over\partial t}+{\bf
u}\cdot \nabla q=-\nabla\cdot \biggl\lbrack {D\over C''(q)}\nabla
{\delta J\over \delta q}\biggr\rbrack,
\end{equation}
where $\delta/\delta q$ is the functional derivative and where we have introduced the free energy
\begin{equation}
\label{ne14qg} J\lbrack q\rbrack=-\int C(q)d^{2} {\bf r}-{1\over
2}\beta\int (q+h)\psi d^{2} {\bf r}.
\end{equation}
We shall now consider the limit of small deformation radius
$R\rightarrow 0$ leading to short-range interactions.  This limit is
appropriate to jovian vortices such as the great red spot (GRS). In
the limit of small deformation radius $R\rightarrow 0$, the relation
(\ref{e2}) between $q$ and $\psi$ can be expanded to second order in
$R^2$ as
\begin{equation}
\label{psiq}\psi\simeq R^{2}(q+h)+R^{4}\Delta q.
\end{equation}
To simplify the discussion further, we shall assume that $C(q)$ is
an even function of $\phi\equiv q-B$, where $B$ is a constant.
Then, the free energy (\ref{ne14qg}) can be written
\begin{equation}
\label{ne17} J\lbrack \phi\rbrack=\beta R^2 \int \biggl\lbrace
{1\over 2}R^2 (\nabla \phi)^2 +V(\phi)-\phi h\biggr\rbrace d^{2}
{\bf r},
\end{equation}
with $V(\phi)=-C(\phi)/\beta R^2 -(1/2)\phi^2$ (we have used the
freedom to introduce a term $\lambda q$ in the free energy to
eliminate the constant $B$). In this limit of short-range
interactions, the free energy is equivalent to the Landau functional
(times $\beta R^2$). Since $\beta$ is negative in cases of physical
interest, we have to minimize the Landau functional.  Note that for
short range interactions, the ensembles are equivalent so that we can
directly work with the free energy functional (\ref{ne17}) instead of the
entropy. In addition, the relaxation equation (\ref{ax1}) becomes
similar to the Cahn-Hilliard equation
\begin{equation}
\label{ne18} {\partial {\phi}\over\partial t}=\beta R^2
\nabla\cdot \biggl\lbrace {D\over C''(\phi)}\nabla (R^2\Delta
\phi-V'(\phi)+h)\biggr\rbrace,
\end{equation}
which appears in the theory of phase ordering kinetics
\cite{bray}. Therefore, Eq. (\ref{ax1}) can be viewed as a generalization of the Cahn-Hilliard equation for long-range interactions \cite{next2003}.

We shall now assume that the potential $V(\phi)$ is a symmetric
function of $\phi$ with two minima at $\pm u$ such that $V(\pm
u)=0$. This is a typical prediction of the statistical theory in the
two-levels approximation where $C(\phi)=(1/2)\lbrace (1+\phi)\ln
(1+\phi)+(1-\phi)\ln (1-\phi)\rbrace$ after normalization
\cite{bs}. However, our goal here is to construct more general
solutions of the Q.G. equations enjoying the same properties. The
potential $V(\phi)$ will have two minima if the equation $C'(u)=-R^2
\beta u$ has non zero solutions. For typical situations, this will be
the case if $\beta>\beta_c =-C''(0)/R^2$. The existence of a critical
(negative) temperature is of course reminiscent of a second order
phase transition in thermodynamics. Note that the functional
(\ref{e18q}) considered by Turkington and collaborators does not
satisfy this condition so that the limit of small Rossby radius
$R\rightarrow 0$ is different in that case and does not lead to a thin
jet structure. 

From now on, we assume that
$\beta>\beta_c$. The stationary solutions of the
Cahn-Hilliard equation (\ref{ne18}) satisfy the differential equation
\begin{equation}
\label{ne19} R^2\Delta \phi=V'(\phi)-h+\alpha,
\end{equation}
where $\alpha$ is an integration constant. The same equation is
obtained by minimizing the Landau free energy (\ref{ne17})
at fixed $\Gamma$. For $R\rightarrow 0$, these solutions describe
``domain walls'' in the theory of phase ordering kinetics. They
connect regions of uniform potential vorticity $q=B\pm u$ separated by
a wall. This is precisely the structure of Jupiter's great red spot
where the PV gradient is concentrated in a thin annular jet scaling
with the Rossby radius.  This analogy has been developed in
\cite{nore,bs} and it was shown to give a good agreement with the
GRS. The present approach makes a connexion with the theory of phase
ordering kinetics and Cahn-Hilliard equations and is valid for an
arbitrary $W$-shape potential $V(q)$.  On a technical point of view,
we differ from \cite{bs} by treating the free energy as a functional
of $q$ (instead of $\psi$), using the short-range expansion
(\ref{psiq}) of $\psi$.  This makes a close link between the free
energy (\ref{ne14qg}) and the Landau free energy (\ref{ne17}). Having
established this link, we can now directly use standard results of
domain wall theory \cite{bray} to describe the great red
spot. Treating the topography as a perturbation of order $R$, we find
that the wall profile is the solution of the equation
\begin{equation}
\label{ne20} R^2 {d^2 \phi\over d\xi^2}=V'(\phi),
\end{equation}
with boundary conditions $\phi(\pm\infty)=\pm u$, where $\xi$ is a
coordinate normal to the wall. This is like the equation of motion
for a fiducial particle in a potential $V$. The first integral is
$R {d \phi/d\xi}=\sqrt{2V(\phi)}$. This result can be used to give
the surface tension
\begin{equation}
\label{ne21} \sigma=R^2\int_{-\infty}^{+\infty}\biggl ({d\phi\over
d\xi}\biggr )^2 d\xi=R\int_{-u}^{+u}\sqrt{2V(\phi)}d\phi.
\end{equation}
Finally, linearizing Eq. (\ref{ne20}) around $\phi=\pm u$ gives
$u\mp \phi\sim {\rm exp}[-\sqrt{V''(u)}\xi/R]$ so that the typical
extension of the PV wall is $L=R/\sqrt{V''(u)}$. The domain wall
curvature due to the topography $h(y)$ can be obtained as follows.
Close to the interface, $\nabla \phi=d\phi/d\xi {\bf n}$ where
${\bf n}$ is a unit vector normal to the wall. Introducing the
curvature $K=\nabla\cdot {\bf n}$, we get $\Delta \phi=K
d\phi/d\xi+d^2 \phi/d\xi^2$. Therefore, Eq. (\ref{ne19}) becomes
to first order
\begin{equation}
\label{ne22} R^2 \biggl (K{d\phi\over d\xi}+{d^2 \phi\over
d\xi^2}\biggr )=V'(\phi)-h(y)+\alpha_1.
\end{equation}
Multiplying by $d\phi/d\xi$ and integrating over the wall, we
obtain the ``curvature-topography'' relation
\begin{equation}
\label{ne23} {K\sigma\over 2u}=\alpha_1-h(y).
\end{equation}
This relation determines the elongation of the GRS under the effect of
an underlying topography (more precisely a deep shear layer). The
above equations are similar to those derived by Bouchet \& Sommeria
\cite{bs} except that they apply to the PV wall instead of the
velocity jet. They are also expressed in the case of an arbitrary
potential $V(\phi)$ so that they can be used to construct a wider
class of models. By construction, all these models are granted to be
nonlinearly dynamically stable via the Q.G. equations. It has already
been shown by Bouchet \& Sommeria
\cite{bs} that this description gives a fair agreement with the
structure of jovian vortices when $S\lbrack q\rbrack$ is the
Fermi-Dirac entropy, corresponding to the two-levels approximation of
the statistical theory. However, some discrepencies have also been
noted (in particular, the predicted surrounded shear is $3$ times
smaller than its real value). It would be interesting to determine
whether an equally good (or even better) agreement can be achieved by
other stationary solutions of the Q.G. equations, that do not
necessarily correspond to statistical equilibria. This could be used
to test the power of prediction of the statistical theory and
determine a class of relevant functions $C(q)$ for jovian flows. The
present paper has given the theoretical tools for constructing
numerically such general solutions and this problem will be considered
elsewhere. These methods can also be extended to the more realistic
shallow-water (SW) equations  \cite{shallow}.

\section{Conclusion}
\label{sec_conclusion}

In this paper, we have proposed a parametrization of 2D geophysical
flows in the form of a generalized Fokker-Planck equation and we have
showed the connexion with Cahn-Hilliard equations in the limit of
small Rossby deformation radius, leading to short-ranged
interactions. This equation is associated with a generalized entropy
functional which is fixed by a prior vorticity distribution in the
sense of Ellis {\it et al.} \cite{ellis}.  This equation can thus
describe the evolution of large-scale motion when the small-scale
forcing has established a permanent vorticity distribution. Its domain
of validity is therefore limited close to equilibrium like in
Onsager's linear thermodynamics. However, it may remain of practical
interest even if we are far from equilibrium. The parametrization
depends on a prior vorticity distribution (or on a generalized
entropy) which is unknown in general and which must be adapted to the
situation. Several prior distributions/entropies that fall in the same
``class of equivalence''
\cite{gfp} should give similar results. We have shown, however, that the 
Fermi-Dirac entropy used by Bouchet \& Sommeria \cite{bs} and the
entropy used by Ellis {\it et al.} \cite{ellis} give different results
in the limit of small Rossby radius. They thus belong to different
``classes''. The Fermi-Dirac entropy may be more appropriate for
describing the Great Red Spot but we have proposed that it may
correspond to a prior vorticity distribution created by a small-scale
forcing with two intense peaks rather than to a free evolution of the
system with two vorticity levels as in \cite{bs}.  These results can
be extended to the more realistic shallow-water equations so their
domain of application is wide. The challenge now amounts to finding
relevant forms of prior vorticity distribution for each specific
situation. This could be achieved by writing stochastic processes for
the generation of potential vorticity (in preparation).

Of course, this thermodynamical parametrization assumes that the
statistical theory works well. However, it has been realized in many
occasions that the ergodic hypothesis which sustains the statistical
theory is not fulfilled everywhere so that the statistical
prediction (assuming efficient mixing) is not truly reliable. Indeed,
the system can be trapped in a stationary solution of the 2D Euler
equation which is not the most mixed state
\cite{staquet,hd,jfm2,brands}. In general, 2D vortices (and
galaxies in astrophysics) are more confined than predicted by the
statistical mechanics of violent relaxation \cite{houches}. The system
tends toward the statistical equilibrium state during violent
relaxation but cannot attain it: the fluctuations die away before the
relaxation process is complete. This phenomenon is referred to as {\it
incomplete relaxation} \cite{lb}. It is an important obstacle for the
general application of the statistical mechanics of violent
relaxation.

This effect of incomplete relaxation can be taken into account in the
relaxation equations by using a space dependent diffusion coefficient of
the form (\ref{e18}) which is related to the local fluctuations of the
vorticity \cite{rr,csr}. This can freeze the system is a sub-domain of
space, in a sort of ``maximum entropy bubble''
\cite{jfm2} surrounded by an un-mixed region $q\simeq 0$ which
is poorly sampled by the flow. This justifies dynamically why the
statistical equilibrium state is not always reached in practice. This
is an interesting property of the relaxation equations which cannot be
obtained with the Turkington-Whitaker algorithm \cite{tw} as it
assumes complete relaxation.  Alternatively, we can try to construct
nonlinearly dynamically stable stationary solutions of the Euler
equations that correspond to confined structures. In that case, the
relaxation equation (\ref{ax1}) can be used as a numerical algorithm
to construct solutions that are not catched by the statistical theory.
In that context, the functional $S[q]$ is interpreted as a
H-function. For example, Tsallis functional is a particular H-function
leading to ``polytropic vortices'' with a compact support (the
enstrophy is a particular case corresponding to $q=2$) \cite{gfp}. Their
confinement is interpreted as an effect of incomplete relaxation due
to lack of mixing/ergodicity. In case of incomplete relaxation, the
goal is not to {\it predict} the metaequilibrium state (this is 
probably an impossible task) but simply to devise a general method for
constructing robust stationary solutions of the 2D Euler equations in
an attempt to {\it reproduce} observed phenomena. This is already a
non-trivial problem by itself. Of course, the statistical theory can
be a valuable {guide} to select physically motivated solutions but the
relaxation equation (\ref{ax1}) can be used to depart from the
statistical prediction and construct a larger class of models. The
same ideas can be developed for collisionless stellar systems and
other systems with long-range interactions \cite{houches}.

The relaxation equation (\ref{ax1}) providing a parametrization of
geophysical flows has been justified from a method similar to
Onsager's linear thermodynamics or from a Maximum Entropy Production
Principle \cite{rsmepp,gfp}. It would be nice to have a more rigorous
justification from ``first principles'' by using kinetic theories. An
attempt has been made in \cite{prl,kin} using formal methods. However,
the domain of application of these methods is limited and does not
correspond to the very nonlinear stages of ``violent relaxation'' that
we would like to describe in concrete situations. This is why more
phenomenological approaches are prefered until we have a more
satisfying theory. Still, the kinetic theory tends to confirm the
general {\it structure} (drift-diffusion) of the relaxation equations
obtained with the MEPP. Finally, we note that generalized
Fokker-Planck equations of the form (\ref{ax1}) appear in other
domains of physics (porous media, Brownian particles in
interaction,...) and biology (chemotactic aggregation of bacterial
populations) where they have a different status and a different
interpretation \cite{gfp,crrs}. Therefore, the general study of these
equations is of considerable interest, independently of the context of
2D hydrodynamics. A systematic study of this class of equations has
been undertaken in \cite{coll}.

\appendix

\section{Relative entropy}
\label{sec_relat}

In this Appendix, we show that the relative entropy (\ref{ngr}) can be seen as a Legendre transform of the mixing entropy (\ref{e9}) when the constraints on $\Gamma_{n>1}$ are treated canonically. The variational principle leading to the Gibbs state  (\ref{e10}) can be written
\begin{equation}
\label{relat1} \delta S-\sum_{n>1}\alpha_{n}\delta\Gamma_{n}-\beta\delta E-\alpha\delta\Gamma-\int \zeta({\bf r})\delta\biggl (\int\rho d\sigma\biggr )d^{2}{\bf r}=0,
\end{equation}
where we have distinguished the robust constraints $E$, $\Gamma$ and
the fragile constraints $\Gamma_{n>1}$. In the point of view of
\cite{miller,rs}, the moments $\Gamma_{n>1}$ are treated
microcanonically and we must ultimately relate the Lagrange
multipliers $\alpha_{n}$ to the constraints. In the point of view of
\cite{ellis}, these constraints are treated canonically by {\it
fixing} the Lagrange multipliers $\alpha_{n}$. If we regard the PV
levels as different species of particles, this is equivalent to fixing
the chemical potentials instead of the total number of particles in
each species.  We are led therefore to define a relative entropy
\begin{equation}
\label{relat2} S_{\chi}= S-\sum_{n>1}\alpha_{n}\Gamma_{n}.
\end{equation}
This is similar to the Legendre transform $F=E-TS$ in usual
thermodynamics when we pass from the entropy (in a microcanonical
description where the energy is fixed) to the free energy (in a
canonical description where the temperature is fixed). Explicitly,
\begin{equation}
\label{relat3} S_{\chi}= -\int \rho\biggl\lbrack \ln \rho+\sum_{n>1}\alpha_{n}\sigma^{n}\biggr\rbrack d^{2}{\bf r}d\sigma.
\end{equation}
Introducing the prior vorticity distribution 
\begin{equation}
\label{relat4} \chi(\sigma)=e^{-\sum_{n>1}\alpha_{n}\sigma^{n}},
\end{equation}
we get
\begin{equation}
\label{relat5} S_{\chi}= -\int \rho \ln\biggl\lbrack {\rho\over \chi(\sigma)}\biggr\rbrack  d^{2}{\bf r}d\sigma,
\end{equation}
which coincides with Eq. (\ref{ngr}) if we set $\mu=\rho/\chi(\sigma)$.

\section{Diffusion coefficient in the Q.G. model}
\label{sec_diffcoeff}

In this Appendix, we derive an expression for the diffusion
coefficient appearing in the parametrization (\ref{ax1}) related to
the Q.G. model. We extend the arguments developed by Robert \& Rosier
\cite{rr} and Chavanis {\it et al.} \cite{csr} (see also \cite{prl}) 
for the 2D Euler equation. The diffusion coefficient can be estimated
by the formula
\begin{equation}
\label{a1} D={1\over 4}\tau\overline{\tilde{u}^{2}}({\bf r},t)
\end{equation}
where $\tau$ is the decorrelation time of the system. Equation (\ref{a1})
corresponds to the general Taylor expression of the turbulent
viscosity. In the Q.G. model, the velocity is related to the potential 
vorticity by 
\begin{equation}
\label{a2} {\bf u}({\bf r},t)=\int q({\bf r}',t){\bf K}({\bf r}-{\bf r}')d^{2}{\bf r}'
\end{equation}
with
\begin{equation}
\label{a3} {\bf K}({\bf r}-{\bf r}')={\bf z}\times {1\over 2\pi R}K_{1}\biggl ({|{\bf r}-{\bf r}'|\over R}\biggr ){{\bf r}-{\bf r}'\over |{\bf r}-{\bf r}'|}.
\end{equation}
where  $K_{1}(x)=-K_{0}'(x)$ is a modified Bessel function. Therefore, the fluctuations of velocity are induced by the fluctuations of potential vorticity so that the expression of the diffusion coefficient becomes
\begin{equation}
\label{a4} D={1\over 4}\tau \int {\bf K}({\bf r}-{\bf r}')\cdot {\bf K}({\bf r}-{\bf r}'') \overline{\tilde{q}({\bf r}',t)\tilde{q}({\bf r}'',t)}d^{2}{\bf r}'d^{2}{\bf r}''.
\end{equation}
We shall now consider that the scale of the spatial correlations $\epsilon$ is smaller than the Rossby radius $R$ and write the correlation function as
\begin{equation}
\label{a5}  \overline{\tilde{q}({\bf r}',t)\tilde{q}({\bf r}'',t)}=\epsilon^{2}\overline{\tilde{q}^{2}}({\bf r}',t)\delta({\bf r}'-{\bf r}'').
\end{equation}
Substituting this expression in Eq. (\ref{a4}), we obtain
\begin{equation}
\label{a6} D={\tau\epsilon^{2}\over 16\pi^{2}R^{2}}\int {K}_{1}^{2}\biggl ({|{\bf r}-{\bf r}'|\over R}\biggr )\overline{\tilde{q}^{2}}({\bf r}',t)d^{2}{\bf r}'.
\end{equation}
Making a local approximation, we can rewrite the foregoing expression in the form
\begin{equation}
\label{a7} D={\tau\epsilon^{2}\over 8\pi} q_{2}({\bf r},t)\int_{\epsilon/R}^{+\infty} {K}_{1}^{2}(\xi)\xi d\xi,
\end{equation}
where $q_{2}=\overline{\tilde{q}^{2}}=\overline{(q-\overline{q})^{2}}$ is the local centered PV enstrophy. The integral can be approximated by $\ln (R/\epsilon)$ so we obtain
\begin{equation}
\label{a8} D={\tau\epsilon^{2}\over 8\pi}\ln\biggl ({R\over\epsilon}\biggr ) q_{2}({\bf r},t).
\end{equation}
We see that the introduction of a finite Rossby radius regularizes the
logarithmic divergence at large scales encountered in the case of the
2D Euler equation \cite{rr,csr,prl}. In the present case, the integral has
to be cut-off at $R$, the typical range of the interactions, while in
the former case it had to be cut-off at the typical vortex or system
size. Finally, the decorrelation time can be estimated by $\tau\sim
\epsilon^{2}/D$. This yields
\begin{equation}
\label{a9} D=K\epsilon^{2}q_{2}^{1/2},
\end{equation}
where $K=\lbrack {1\over 8\pi}\ln (R/\epsilon)\rbrack^{1/2}$ is a constant of order unity.

\section{Non-equilibrium distributions}
\label{sec_noneq}

The relaxation equation (\ref{ax1}) gives the evolution of the coarse-grained PV $\overline{q}({\bf r},t)$ out of equilibrium. This evolution respects the conservation of energy and circulation and depends on the prior $\chi(\sigma)$ through the function $C(q)$. We may wonder about the expression of the detailed PV distribution $\rho({\bf r},\sigma,t)$ out of equilibrium. One idea is to determine this distribution by using a maximum entropy principle. Specifically, one may argue that  $\rho({\bf r},\sigma,t)$ maximizes the relative entropy (\ref{ngr}) at fixed coarse-grained PV $\overline{q}({\bf r},t)=\int \rho\sigma d\sigma$ and normalization $\int \rho d\sigma=1$. This yields  
\begin{equation}
\label{noneq1} \rho({\bf r},\sigma,t)={1\over Z({\bf r},t)}\chi(\sigma)e^{-\sigma\Phi({\bf r},t)},
\end{equation}
where $Z$ and $\Phi$ are Lagrange multipliers determined by
\begin{equation}
\label{noneq2} Z({\bf r},t)=\int \chi(\sigma)e^{-\sigma\Phi({\bf r},t)}d\sigma,
\end{equation}
\begin{equation}
\label{noneq3} \overline{q}({\bf r},t)={1\over Z({\bf r},t)}\int \chi(\sigma)\sigma e^{-\sigma\Phi({\bf r},t)}d\sigma.
\end{equation}
This is similar to the equilibrium problem with $\Phi({\bf r},t)$ instead of $\beta\psi+\alpha$. In particular, the relation (\ref{noneq3}) can be rewritten
\begin{equation}
\label{noneq4} \overline{q}({\bf r},t)=F\lbrack \Phi({\bf r},t)\rbrack,
\end{equation}
where $F$ is the same function as in Eq. (\ref{e11}), entirely determined by the prior $\chi(\sigma)$. Thus, knowing $\overline{q}({\bf r},t)$ from the evolution equation (\ref{ax1}), we can inverse (at each time) Eq. (\ref{noneq4}) to obtain $\Phi({\bf r},t)$, then 
$\rho({\bf r},\sigma,t)$. On the other hand, from the relations $q'\lbrack \Phi({\bf r},t)\rbrack=-q_{2}({\bf r},t)$ and $C'\lbrack \overline{q}({\bf r},t)\rbrack=-\Phi({\bf r},t)$ leading to $\overline{q}'\lbrack \Phi({\bf r},t)\rbrack=-1/C''\lbrack \overline{q}({\bf r},t)\rbrack$, we get
\begin{equation}
\label{noneq5}q_{2}({\bf r},t)={1\over C''\lbrack \overline{q}({\bf r},t)\rbrack}.
\end{equation}
Therefore, this approach justifies, or at least is consistent, with the use of Eq. (\ref{org}) out of equilibrium.

\end{document}